\documentclass{aastex631}
\usepackage{booktabs}
\usepackage{amsmath}

\usepackage{float}

\begin{document}

\title{Insights from the Gaussian Processes Method for the FRB-associated X-ray Burst of SGR 1935+2154}

\author{Ruijing Tang}
\affiliation{Department of Physical Science and Engineering, Beijing Jiaotong University,
3 Shangyuancun, Haidian District, Beijing 100044, China
}
\author{Dahai Yan}
\email{yandahai@ynu.edu.cn}
\author{Haiyun Zhang}
\affiliation{Department of Astronomy, Key Laboratory of Astroparticle Physics of Yunnan Province, Yunnan University, Kunming 650091, China}

\author{Qingchang Zhao}
\author{Lian Tao}
\email{taolian@ihep.ac.cn}
\author{Chengkui Li}
\author{Mingyu Ge}
\author{Xiaobo Li}
\author{Qianqing Yin}
\affiliation{Key Laboratory of Particle Astrophysics, Institute of High Energy Physics, Chinese Academy of Sciences, Beijing 100049, China}

\author{Ce,Cai}
\affiliation{College of Physics, Hebei Normal University, Shijiazhuang, Hebei 050024, China}



\begin{abstract}
Gaussian processes method is employed to analyze the light curves of bursts detected by \textit{Insight}-HXMT, \textit{NICER}, and \textit{GECAM} from SGR 1935+2154 between 2020 to 2022. It is found that a stochastically driven damped simple harmonic oscillator (SHO) is necessary to capture the characteristics of the X-ray bursts. Variability timescale of the X-ray bursts, corresponding to the broken frequencies in the SHO power spectral densities (PSDs), are extracted. In particular, a high broken frequency of 35 Hz where the index of the SHO PSD changes from $-4$ to $-2$ is constrained by the HXMT-HE burst associated with FRB 200428. It is suggested that the corresponding timescale of 0.03\,s could be the retarding timescale of the system driven by some energy release, and the production of the HE photon should be quasi-simultaneous with the response. The other special event is a \textit{NICER} burst with a retarding timescale of $1/(39\ {\rm Hz})\approx0.02$ s. In the normal X-ray bursts, no retarding timescale is constrained; a long relax/equilibrium timescale (corresponding to a broken frequency of 1--10 Hz where the index of the SHO PSD changing from $-4$/$-2$ to 0 in the SHO PSD) is obtained. The results indicate that the FRB-associated HXMT-HE X-ray burst could be produced immediately when the system is responding to the energy disturbance, far before the equilibrium state.
\end{abstract}



\section{Introduction} \label{sec:intro}

Magnetars, a special type of neutron star, primarily derive energy from a powerful magnetic field. They were first noticed at the end of the 1970s \citep{mazets1979observations}. The energy of gamma-ray bursts was considered as dissipation of magnetospheric energy from hyper-magnetized neutron stars \citep{1982ApJ...260..371K}. These astronomical objects have been the focus of intense research due to their unique features, including significant X-ray variability, short spinning periods (2--12,s), fast spinning down velocity ($\sim 10^{-13}-10^{-10}~{\rm s}^{-1}$), and strong dipole fields around $10^{13}-10^{15}$~G \citep{Mereghetti2016}. Based on the characteristics of burst emissions, two categories are commonly identified: Soft gamma-ray repeaters (SGRs) and anomalous X-ray pulsars (AXPs). Distinguishing phenomena of SGRs include intermittent bursts of soft gamma-ray radiation. Rare GRBs associated with SGRs have an impact at least two orders of magnitude higher than relatively smaller events \citep{1982ApJ255L45C,2008A&ARv1225M}.

Magnetars bursts had been detected since March 5, 1979, from SGR 0526–66. Initially, signals were categorized because of their softer spectrum and were generated repeatedly from specific directions in the sky. Therefore, they are called soft gamma-ray repeaters (SGRs) \citep{1992ApJ392L9D,Narayan1992GammarayBA}. The classical model of SGRs suggests that a plausible mechanism triggering short bursts comes from the dramatic activity of crusts, which occurs due to the accumulation of stress in the crust as the strong magnetic field slowly diffuses through the dense stellar matter. When the solid crust is unable to modify plastic deformation to reach the required equilibrium profile constrained by the spin-down process or when the magnetic stress surpasses the shear modulus, crustquakes are likely to occur \citep{BAYM1971816,1995MNRAS275255T}. In subsequent years, the initial magnetar model underwent substantial modification and expansion, primarily driven by the incorporation of multi-wavelength observations. SGRs, as generators of gamma-ray, X-ray, and radio bursts, have the potential to produce Fast Radio Bursts (FRBs) with the highest energy and millisecond duration. Although many models have been proposed \citep[e.g.][]{2018ApJ86831Y,2020ApJ8971L,2020ApJ904L15Ioka,Yang_2021,2021MNRAS.507.2208W}, the generating process of FRBs remains unclear.

SGR J1935+2154 (henceforth, SGR J1935) is a Galactic magnetar. The Burst Alert Telescope (BAT) of the Neil Gehrels Swift Observatory initially detected it in July 2014, with subsequent observations by Chandra and XMM-Newton between 2014 and 2015 \citep{8164910}. The location of source, identified by \textit{Swift}/BAT, is near geometric center of the Galactic supernova remnant (SNR) G57.2+0.8 at a distance of 9.1 kpc \citep{2014GCN.165331G}. SGR J1935 has a rotation period of 3.24\,s ($P=3.24$~s) and a dipole magnetic field of $2\times10^{14}$~G inferred from spin-down characteristics. It experienced an activating period from April 27th, 2020. Hundreds of X-ray bursts were recorded by \textit{INTEGRAL} \citep{2020ApJ898L29M}, \textit{Insight}-HXMT \citep{2021NatAs378Lhxmt}, \textit{NICER} \citep{Younes_2020}, \textit{Swift} \citep{2020GCN.276651Swift}, \textit{AGILE} \citep{Tavani2021}, \textit{Konus-Wind} \citep{2021NatAs5372R} and \textit{NuSTAR} \citep{Borghese_2020}. Surprisingly, one unique fast radio burst (FRB 200428) was detected from it by \textit{CHIME} \citep{2020Natur.58754C} and \textit{STARE2} \citep{2020Natur.58759B} with associated X-ray burst on April 28 UTC 14:34:24. Luckily, \textit{Insight--HXMT} satellite captured this event \citep{2021NatAs378Lhxmt}. Thus, this X-Gamma Burst co-emitting event gave an unprecedented opportunity to figure out the FRB generating mechanism and relationship with X-ray Burst (XRB) further. This source experienced another period of activity in the middle of 2021, prompting various high-energy missions, such as \textit{Fermi} \citep{Lesage2021}, \textit{INTEGRAL} \citep{2020ApJ898L29M}, \textit{GECAM} \citep{Xiao2021}, \textit{Swift} \citep{Palmer2021}, \textit{Konus-Wind} \citep{Ridnaia2021}, and \textit{Calet} \citep{Nakahira2021}. 
Three transition states of radio bursts observed in 2022 are regarded as intermediate classes between FRBs and normal radio bursts categories \citep{2022ATel15681....1D,2022ATel15682....1W,2022ATel15708....1L}. Undoubtedly, they provide valuable evidence for intensive study of burst-generating mechanism, especially following FRB 200428.


For co-emitting XRBs, the total duration ranges from 0.3 to 0.5\,s, with individual sub-burst peaks lasting approximately 10\, ms \citep{2021NatAs378Lhxmt,2020ApJ898L29M, Tavani2021}. The isotropic energy is estimated to be $(0.5-1.2)\times 10^{40}$ erg (for a distance of 10 kpc), which is about $10^{5}$ times larger than that observed in the radio band \citep{2021NatAs378Lhxmt}. In comparison, FRB signals have an overall duration of around 30\,ms, with two sub-bursts each lasting approximately 5\,ms. The isotropic radiated energy from the FRB signal is estimated to be in the range of $(0.3-2.4) \times  10^{35}$ erg \citep{Yang_2021, Yuan_2020}. A significant glitch was observed in SGR J1935, occurring around $3.1 \pm 2.5$ days before the emission of FRB 200428. The glitch involved changes in frequency ($\Delta\nu = 19.8 \pm 1.4$ µHz) and spin-down rate ($\Delta\dot{\nu} = 6.3 \pm 1.1$ pHz s$^{-1}$). The associated change in spin-down power rate, $\Delta \dot{\nu}/\dot{\nu}$, ranks among the highest observed in pulsar glitches, which includes a delayed spin-up process \citep{ge2022giant}. There are indications of temporal spectral hardening linked to the occurrence of two peaks in the X-ray burst \citep{2021NatAs378Lhxmt,2020ApJ898L29M}. The inherent temporal gap between X-ray and radio emission peaks is estimated to be within approximately 10 ms \citep{Yamasaki_2022}.

In this paper, a comprehensive analysis is conducted that focuses on analyzing the unique properties of burst light curves using Gaussian Processes. We employ Gaussian Processes to fit light curves and extract physical insights from timing properties of bursts. 
The subsequent sections are structured as follows: 
Section \ref{sec:style} presents observations and data reduction, from \textit{Insight}-HXMT, \textit{NICER}, and \textit{GECAM}. 
Section \ref{sec:analysis} briefly introduces the Gaussian Processes method.
Section \ref{sec:results} demonstrates fitting results.
Section \ref{sec:discussion} proposes a physical scenario to interpret results.

\section{OBSERVATIONS and DATA REDUCATION} \label{sec:style}
\subsection{Insight-HXMT}
\label{HXMT}

\textit{Insight}-HXMT, an X-ray astronomy telescope, aims to observe X-ray sources within wide energy range of 1 to 250 keV. It studies X-ray sources to gain deeper insights into mechanisms and radiation processes in strong magnetic and gravitational fields, and other extreme conditions. Three payloads are involved: High Energy X-ray telescope (HE; 20--250 keV); Medium Energy X-ray telescope (ME; 5--30 keV);  Low Energy X-ray telescope (LE; 1--15 keV). SGR J1935+2154 was observed by \textit{Insight}-HXMT during two periods: the first started from 2020-04-28T07:14:51 UTC to 2020-06-01T00:00:01 UTC, lasting 33 days, and the second began from 2022-10-13T04:51:41 UTC to 2022-11-01T15:48:29 UTC, lasting 19 days. Brief information about the observations is presented on the Insight-HXMT website\footnote{http://archive.hxmt.cn/proposal}. 

The data was processed using the \textit{Insight}-HXMT Data Analysis software (HXMTDAS 2.05). The clean and calibrated events files were extracted based on several criteria: (1) Earth elevation angle larger than $10^{\circ}$, (2) geomagnetic cut-off rigidity larger than 8GV, (3) the pointing offset angle less than $0.04^{\circ}$, (4) at least 300 s before and after the South Atlantic Anomaly passage. The arrival time is corrected using the \texttt{hxbary} tool included in the HXMTDAS software package.


\subsection{NICER}
Neutron Star Interior Composition Explorer (\textit{NICER}), situated on the International Space Station (\textit{ISS}), covers the 0.2--12 keV energy range \citep{Younes_2020}. \textit{NICER} observed this source from 2020-4028T00:40:58 UTC, lasting 60 days. The tools for processing data are NICER Data Analysis Software (\texttt{NICERDAS 2022-01-17\_V009}), \texttt{HeaSoft v 6.30}, and Calibration Database \texttt{CALDB xti20210707}. Using the \texttt{nicerl2}, we produced clean event files with general calibration and screening criteria. The Solar System barycentric correction to the photon arrival times is applied with the \texttt{barycorr} tool. The light curves are extracted by pipeline \texttt{nicerl3$-$lc}.


\subsection{GECAM}\label{GECAM}

Gravitational wave high-energy Electromagnetic Counterpart All-sky Monitor (GECAM) is a constellation designed to monitor gamma-ray bursts \citep{li2022b, xu2022}. Each satellite has two class payloads: gamma-ray detectors (GRD) and Charged Particle Detectors (CPDs). The mission can locate X-ray and gamma-ray bursts from 6keV to 5MeV. GECAM detected two bursts from SGR 1935+2154, which are associated with two transition state radio bursts. They were detected on October 14th, 2022 (FRB 221014A) and November 20th, 2022 (FRB 221120A), respectively. We process the data using the standard pipeline to get light curves for further analysis \citep{song2020}.



\section{Analysis} \label{sec:analysis}
\subsection{Timing analysing models} 
Gaussian Processes is a probabilistic theory to make predictions about a continuous stochastic process, 
namely by defining a distribution over functions \citep[e.g.,][]{2006gpml.book.....R}. 
It is supported by the mean function $m(x)$, and covariance function $k(x, x_{0})$. This represents a logical extension of the Gaussian distribution, where the mean and covariance functions are respectively displayed by vector and matrix. In the astronomy field, specific types of target objects exhibit intricate internal characteristics in observatory data that match the physical distribution in the Gaussian Process. Therefore, it is favored in data analysis \citep{Foreman-Mackey_2017}, widely used to fit light curves of various astronomical objects and events \citep{aigrain2022gaussian}.
It typically serves for the following cases: gravitational lensed quasar, transit, and active galactic nuclei \citep[e.g.,][]{2022ApJ...930..157Z,2023ApJ...944..103Z}. 

SGR J1935+2145 yields consecutive bursts spanning multiple energy bands in its active states. 
We use the {\it celerite} package in Python, involving a general-defined kernel function as an essential part of the modeling process \citep{Foreman-Mackey_2017}. The kernel function is a covariance function $k_{\alpha}(X_{n},X_{m})$, which depicts the potential correlation between points \begin{equation}
    k_{\alpha}(t_{nm})=\sigma_{n}^{2}\delta_{nm}+\sum_{j=1}^{J}a_{j} e^{-c_{j}t_{nm}}
\end{equation} and when importing quasi-periodic terms containing oscillation process,
\small
\begin{equation}
    k_{\alpha}(t_{nm}) = \sigma_{n}^2\delta_{nm} + \sum_{j=1}^{J} \left( a_{j}e^{-c_{j}t_{nm}}\cos(d_{j}t_{nm})+ b_{j}e^{-c_{j}t_{nm}}\sin(d_{j}t_{nm}) \right)
\end{equation}
\normalsize Involving complex terms and giving zero value for $b_{j},d_{j}$ parameters, simplify formula appears $k_{j}(t_{nm})=a_{j}e^{-c_{j}t_{nm}}$, which also called "damped random walk" (DRW) \citep{1992ApJ398169R}. 
The DRW model is used to depict a disturbed system returning to equilibrium due to the damping effect. It is generated from the Langevin equation, \begin{equation}
    \left[\frac{d}{dt}+\frac{1}{t_{\rm DRW}}\right]y(t)=\sigma_{\rm DRW}\epsilon(t)
\end{equation} After applying the Fourier transform, the PSD is obtained
\begin{equation}
    S_{j}(\omega)=\sqrt{\frac{2}{\pi}}\frac{a_{j}}{c_{j}\times(1+(\frac{\omega}{c_{j}})^2)}
    \label{psd4}
\end{equation}

The stochastically driven damped simple harmonic oscillator (SHO) 
is more complex,
and the DRW model resembles a specific case in SHO with specific parameter conditions. 
The SHO equation is written as
\begin{equation}\label{5}
    \left[\frac{d^2}{dt^2}+\frac{\omega_{0}d}{Q dt}+\omega_{0}^2\right]y(t)=\epsilon(t),
\end{equation} 
and meaning of parameters appear in Equation \eqref{5}: the frequency of undamped oscillation system $\omega_{0}$, quality factor $Q$, 
and random driving force $\epsilon(t)$. Setting stochastic driving force to be white noise, PSD function is \begin{equation}
    S(\omega)=\sqrt{\frac{2}{\pi}}\frac{S_{0}\omega_{0}^4}{(\omega^2-\omega_{0}^2)^2+\frac{\omega_{0}^2\omega^2}{Q^2}}
    \label{psd6}
\end{equation}
which comes from Fourier transform of auto-covariance function (ACVF).
The value of $(S_0)$ is directly proportional to power at frequency $w = w_0$ with function
$S(\omega_{0})=\sqrt{2/\pi}S_{0}Q^2$ \citep{1990ApJ364699A}. The fundamental kernel function is presented in below and $f(t, Q,\omega_{0})$ varies when Q is in specific range \citep{Foreman-Mackey_2017}.
\begin{equation}
            k_{\rm SHO}(t,S_{0},Q, \omega_{0}) = S_{0}\omega_{0}Q e^{-\frac{\omega_{0} t}{2Q}} 
 f (t, Q,\omega_{0})
\end{equation}
According to the $Q$ value, three modes are classified with corresponding damping states. 
To be specific, 
when $Q$ is much smaller than 0.5, i.e., indicating the overdamped region,
 SHO model resembles closely DRW model; 
when $Q\sim0.5$, the oscillation is in the critical damped region;
lastly, when $Q$ is larger than 0.5, i.e., in the underdamped region,
it implies high-quality oscillation and has a similar profile of PSD to Lorentzian \citep[e.g.,][]{Foreman-Mackey_2017,2021ApJ...919...58Z}.

The SHO model can describe more complex behaviors of a varying system. In particular, it can provide information on the stage before the system strongly responds to disturbance, i.e., the $\nu^{-4}$ regime in the SHO PSD.

Based on equations (\ref{psd4}) and (\ref{psd6}), 
PSD can be graphed, providing direct insights into the variability from frequency space. 
The general equation for power spectrum is
\begin{equation}
 \rm PSD (\nu)=|\int_{-\infty}^{\infty} ACVF(t)e^{-i\nu t}dt|^2\ ,
    \label{generalpsd}
\end{equation}
implying the decomposition of power within time series into a linear combination of complex harmonics exponentials, 
each figured by amplitude and frequency.
In DRW model, a single slope power law of $\nu^{-2}$ frequency is typically observed at frequencies larger than the $\nu_{\rm break}$ frequency; 
while below $\nu_{\rm break}$, the flat section of the PSD indicates uncorrelated white noise.

Theoretically, the SHO PSD shows a regime with a steeper slope ( $\nu^{-4}$) at high frequencies \citep[e.g.,][]{2017MNRAS.470.3027K,2019PASP..131f3001M}, indicating that
the system responds weakly to high-frequency components of disturbance.
In the $\nu^{-4}$ regime, the behavior of the system is determined by an interaction between inertia of system and disturbance acting on system.
Therefore, there is an additional frequency break 
where the slope of PSD becomes $\nu^{-4}$ from $\nu^{-2}$,
$$\nu_{\rm SHO-break1}=min(\frac{2\pi}{\sqrt{\alpha_{1}^{2}-2\alpha_{2}}}, \frac{2\pi}{\sqrt{\alpha_{2}}}),$$
where $\alpha_{1}=\omega_{0}/Q$, $\alpha_{2}=\omega_{0}^{2}$.
This break frequency can be used to investigate 
the dominated physical process before equilibrium, 
e.g., the timescale of disturbance process or the responding timescale of system.
In $\nu^{-2}$ regime, system responds strongly to the disturbance, 
meanwhile damping effect becomes important.
The transition from the $\nu^{-2}$ regime to the flat regime reflects that system achieves a balance between disturbance and damping effect.

Two roots $r_{1}$ and $r_{2}$ solved from equation (\ref{5}) generate rise and decay timescales $t_{\rm rise}$ and $t_{\rm decay}$, 
corresponding to two broken frequencies in SHO PSD at $\nu^{-4}\mapsto\nu^{-2}$ and  
$\nu^{-2}\mapsto\nu^{0}$, respectively \citep{2019PASP..131f3001M,Nakariakov_2021}. 
\subsection{Fitting Method}
To get the posterior distribution of model parameters, 
Markov Chain Monte Carlo (MCMC) is run for 50,000 steps with a 32-step width each time, and initial 20,000 steps are dropped for proper estimation. 
This fitting procedure is achieved by running the {\it emcee} package \citep{Foreman-Mackey_2013}. 

Standardized residuals and autocorrelation function (ACF) of standardized residuals and squared residuals serve as indicators for assessing fitting quality \citep[e.g.,][]{2014ApJ...788...33K}. The distribution pattern of standardized residuals across both time and numerical scales concretely reflects model's performance. When distribution of values resembles normal distribution with a peak superimposing zero value, 
it is considered to exhibit the characteristics of a well-fitting model. 
The stochastic fluctuation of ACF values within 95\% confidence interval boundaries of white noise is positive indication that model  
captures all structure in data \citep[e.g.,][]{2014ApJ...788...33K}. 
The AIC (Akaike Information Criterion) test indicates fitting quality. Smaller values gained from AIC suggest well-fitting, which is prefered.

\section{Results}\label{sec:results}
To ensure data quality while capturing as much detail as possible in XRBs generated from magnetars, we choose the following time bins for different telescopes, i.e., 
setting \textit{NICER} light curves in 4 ms bin, \textit{Insight}-HXMT light curves in 5 ms bin, 
the one associated with fast radio burst in 20 ms, 
and GECAM light curves in a 20 ms bin.
We put those data in Gaussian Processes analysis to investigate variability pattern of the XRBs.
In short, we use 10 light curves from \textit{NICER}, 28 from \textit{Insight}-HXMT, and 2 from \textit{GECAM} for statistical analysis. 

\subsection{Model Selection}
\begin{figure}[h!]
   \centering
    \includegraphics[width=0.40\textwidth]{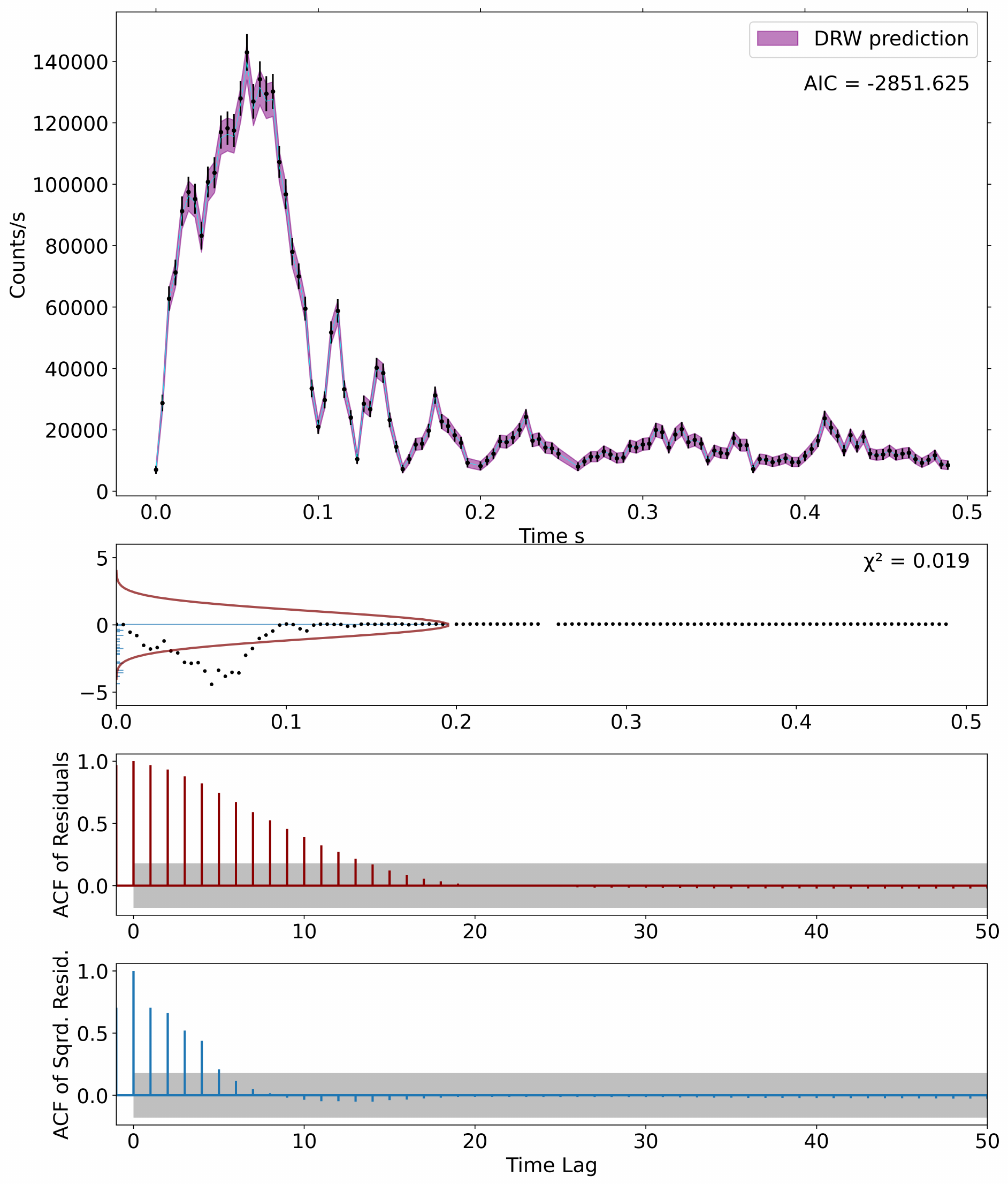} 
    \includegraphics[width=0.40\textwidth]{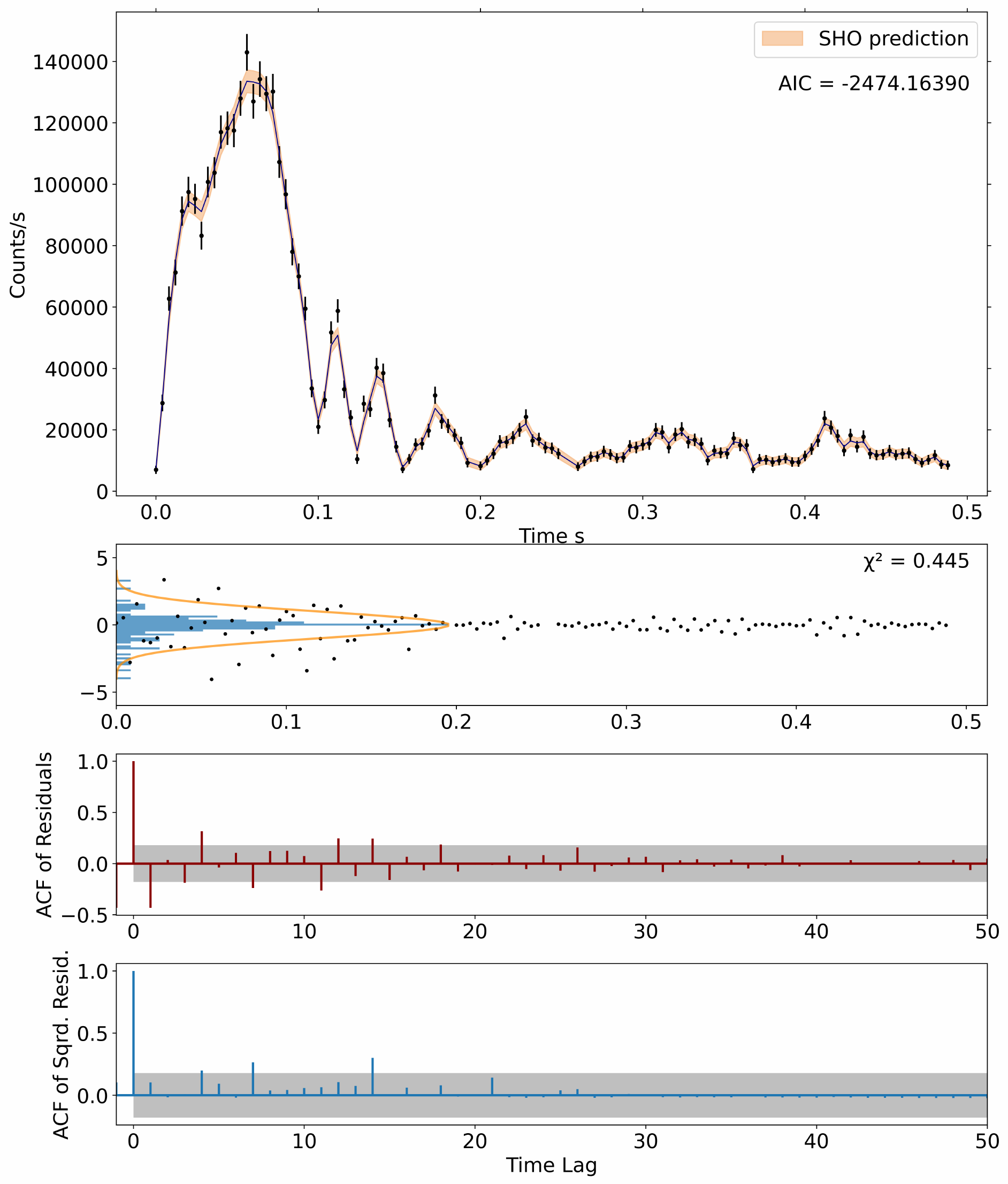}
    \caption{Fitting results for one light curve (\textit{NICER}-1) with the DRW kernel (left panel) and SHO kernel (right panel).
    The distribution of standard residuals (the panel below corresponding light curve) is fitted with Gaussian distribution, and reduced $\chi^2$ is given in panel.
    The autocorrelation functions of residuals and squared residuals (the two bottom panels) are more important to test fitting quality.
    There are clear non-white-noise structures in fitting results with DRW kernel, while there are no such structures in SHO. Therefore, in this case, DRW model is not favored and SHO is preferred to interpret X-ray burst of magnetar.
    }
    \label{sampledrw}
\end{figure}

We apply the two kernels of DRW and SHO to data initially. 
Finally, SHO turns out to have lower AIC values, passing the statistical tests of residuals.
For example, Figure \ref{sampledrw} presents good evidence to verify that DRW is unsuitable, 
where both the autocorrelation functions of residuals and squared residuals clearly exceed 95\% confidence interval boundaries of white noise. This indicates that the DWR kernel cannot capture structures in data.
It is found that SHO delivers good performance in the fittings, and can well describe the structures of light curves from XRBs 
(see the right panel of Figure~\ref{sampledrw} as example).
SHO kernel is preferred over DRW kernel for all light curves we considered here.

\subsection{SHO Results}

\begin{figure*}[h!]
    \centering
    \begin{minipage}[b]{0.325\linewidth}
        \includegraphics[width=\textwidth]{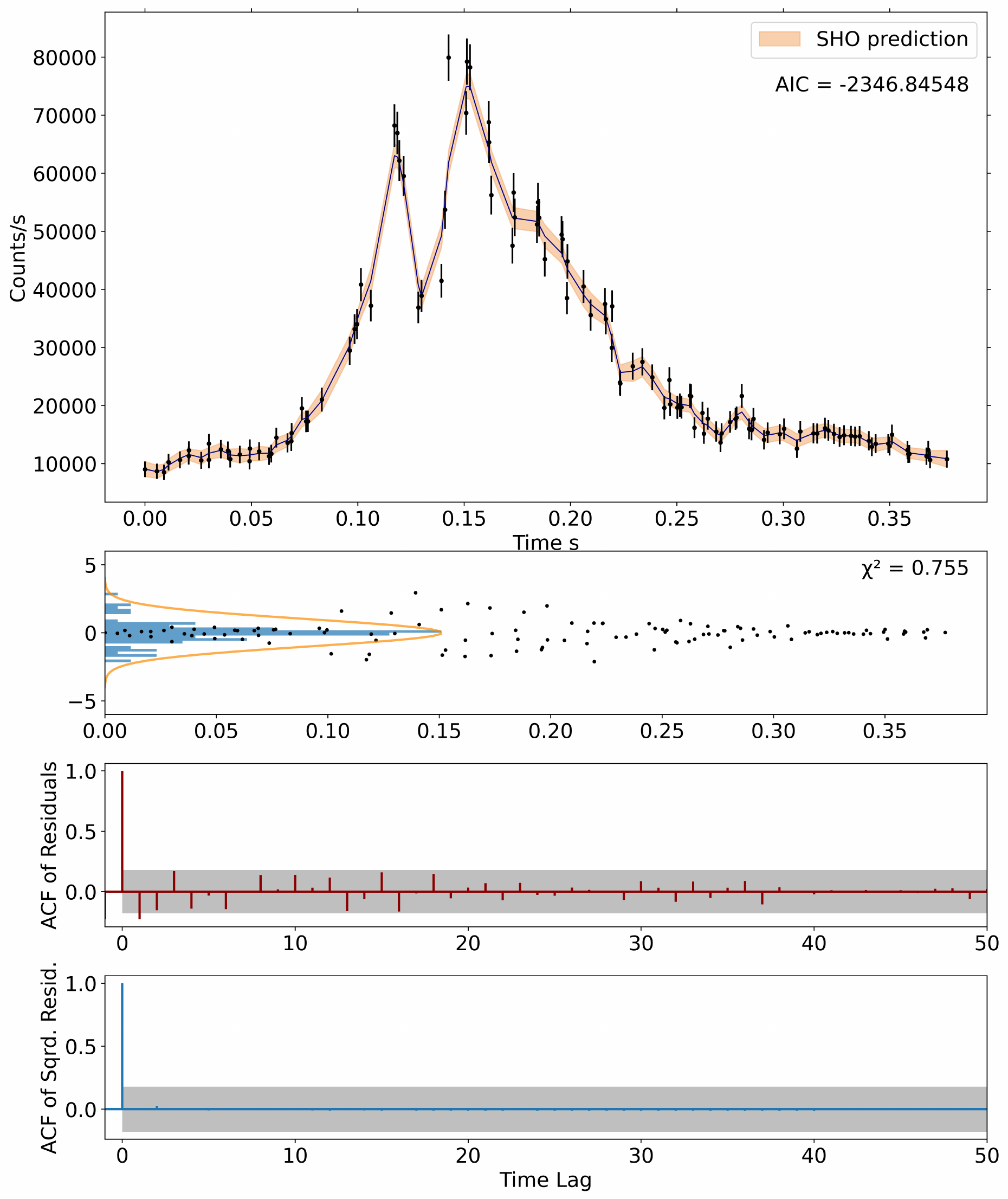}
    \end{minipage}
    \begin{minipage}[b]{0.325\linewidth}
        \includegraphics[width=\textwidth]{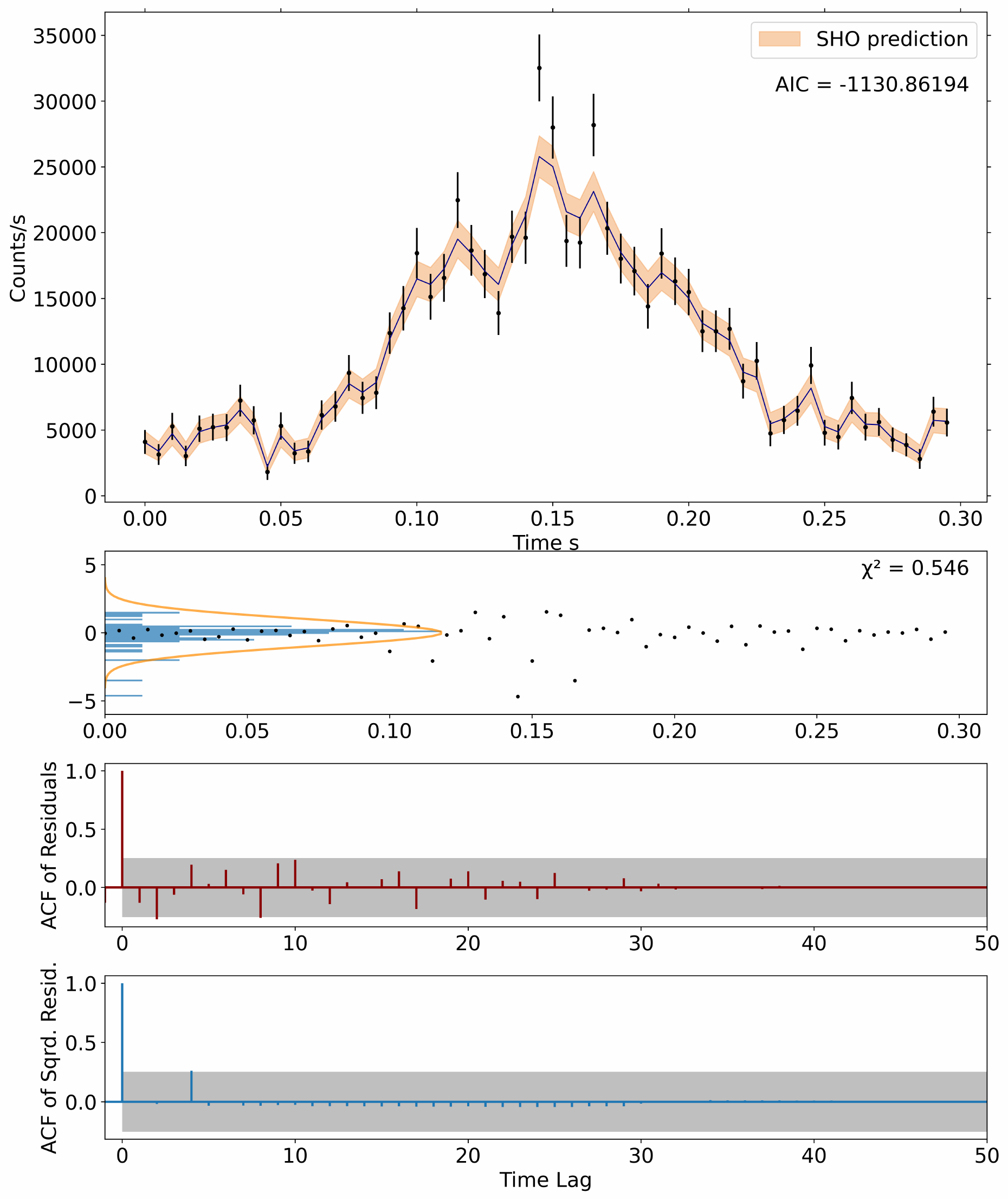}
    \end{minipage}
    \begin{minipage}[b]{0.325\linewidth}
        \includegraphics[width=\textwidth]{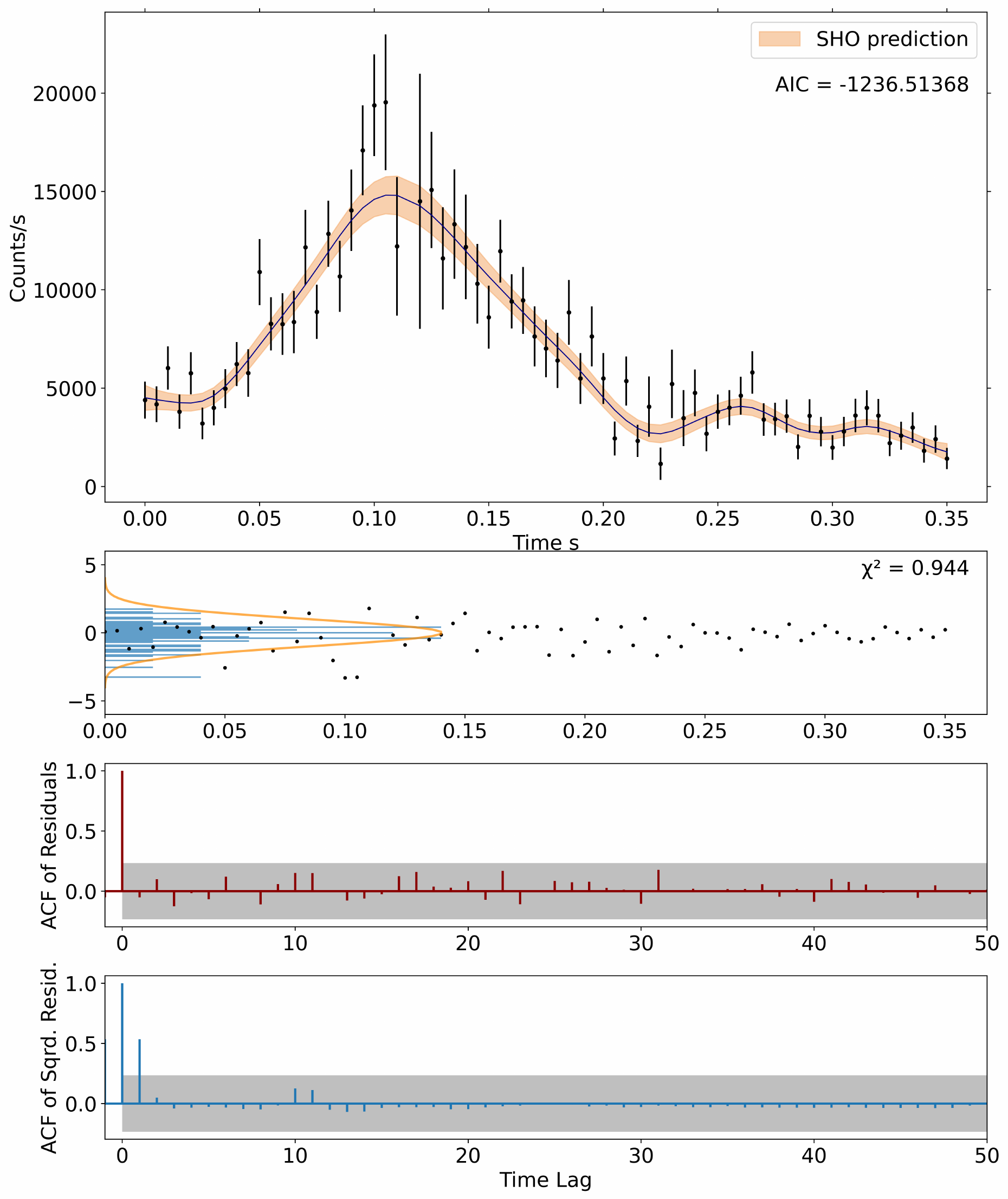}
    \end{minipage}
    \caption{Fitting results for the X-gamma-ray data from \textit{Insight-HXMT} using SHO model.
            These three X-gamma-ray bursts are associated with two radio pulses of FRB 200428.
            The light curves are binned into 5 ms intervals.
            The panels from left to right represent fitting results for HE, ME, and LE respectively. 
            The starting time of HE light curve is defined as at 14:34:24.427 on April 28th, 2020. 
            The  ME light curve segment offsets at 14:34:24.303 on April 28th, 2020, and the LE light curve segment starts at 14:34:24.302. 
            }
    \label{Figurecombin}
\end{figure*}

\begin{figure}
    \centering
    \includegraphics[width=0.32\textwidth]{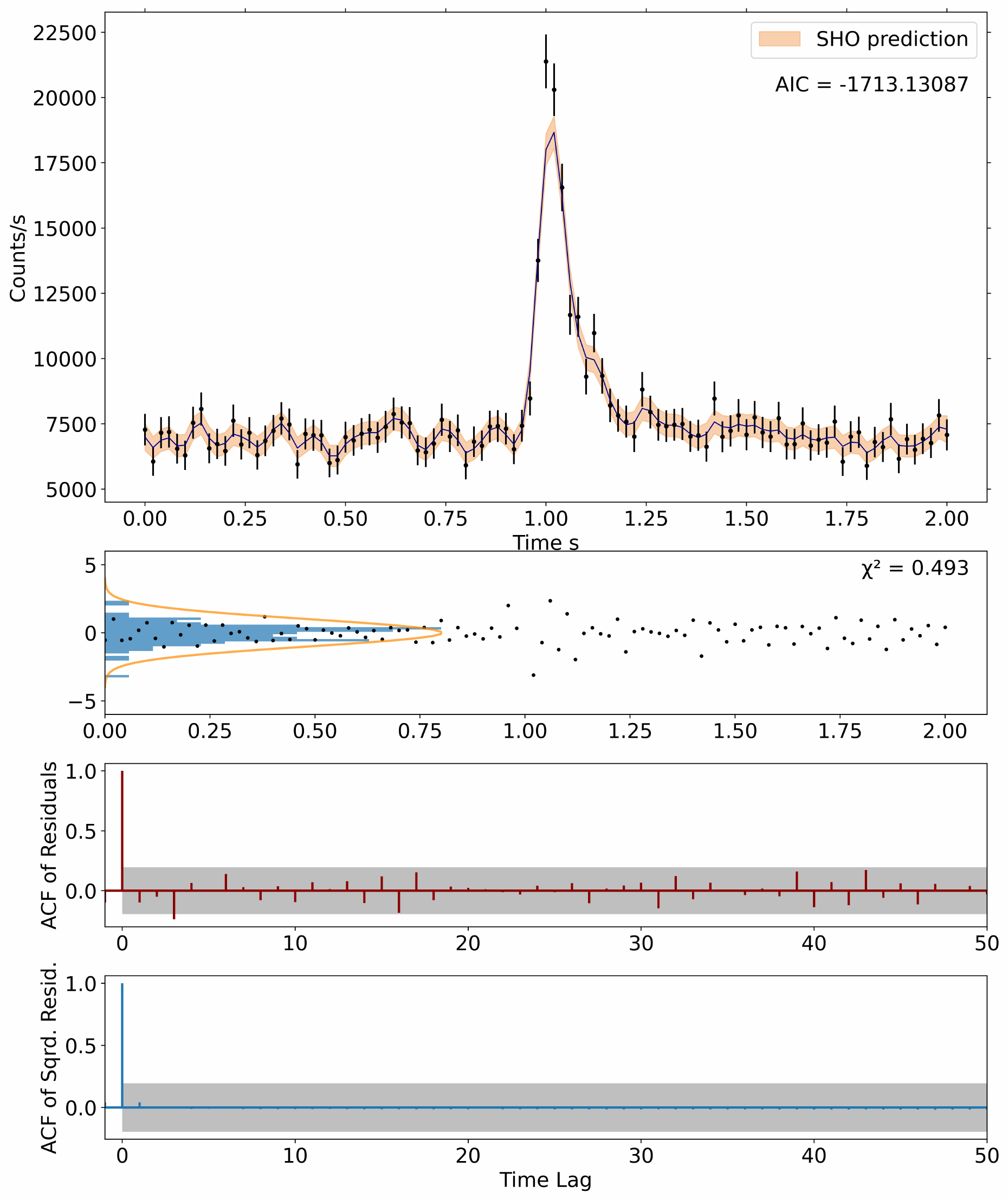}
    \includegraphics[width=0.32\textwidth]{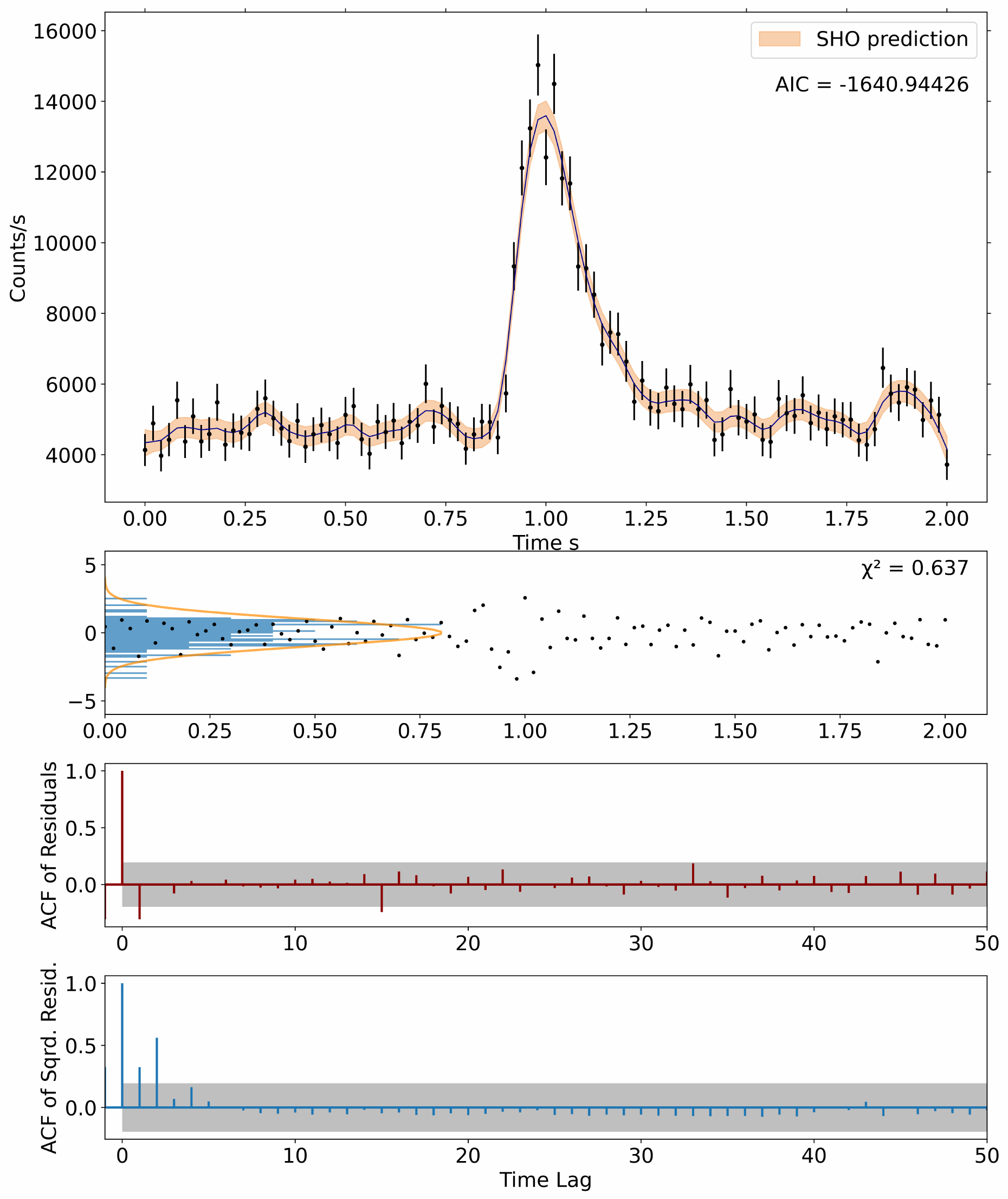}
    \includegraphics[width=0.32\textwidth]{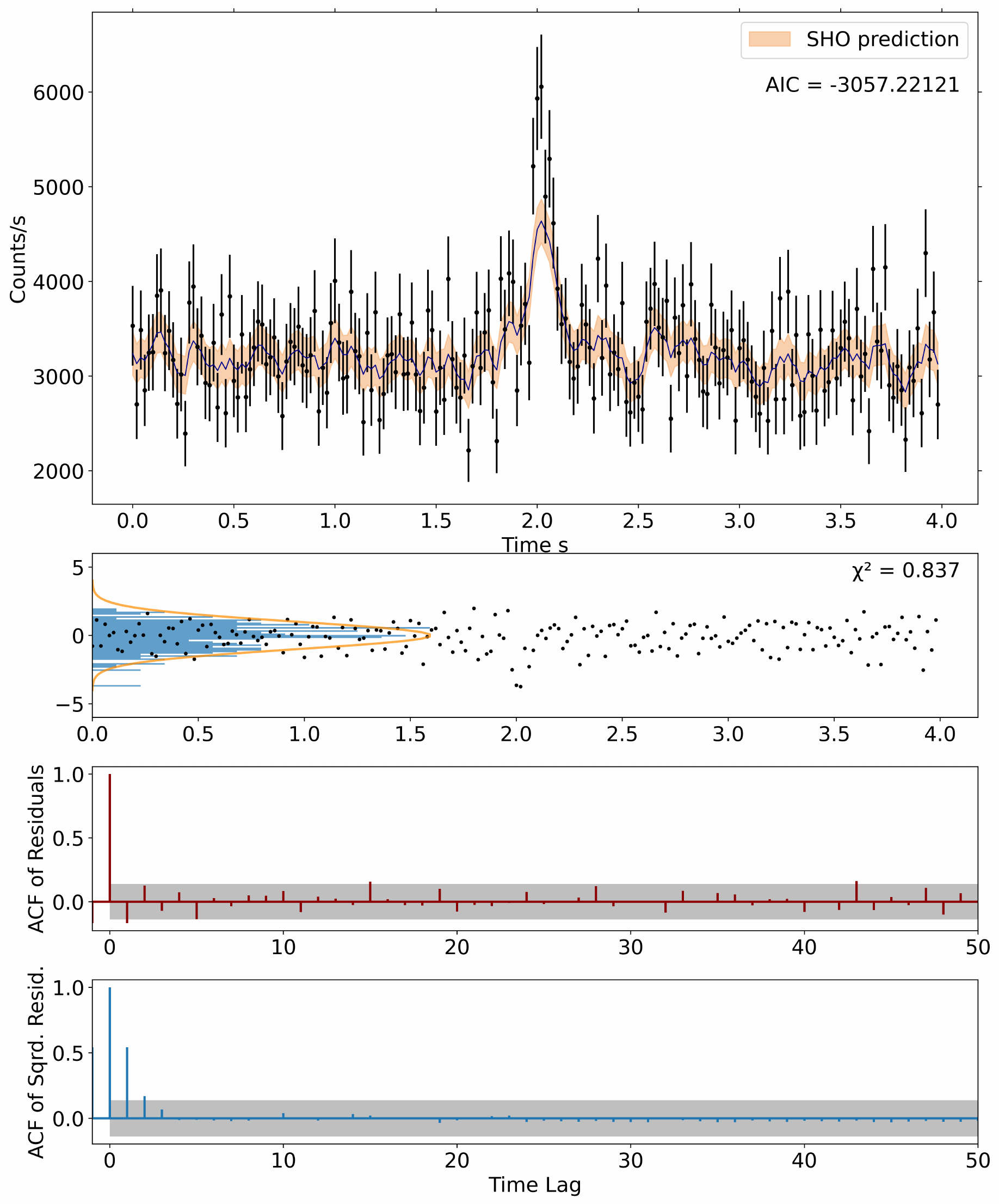}
    \caption{SHO fitting results for three bursts associated with transition state of radio bursts. 
            The first light curve was detected by GECAM on October 14th, 2022, and the middle one was also recorded by GECAM on November 20th. The last light curve was observed by HXMT on October 21st.}
    \label{fig:2022}
\end{figure}

 The two significant X-ray peaks associated with FRB signals have been captured by HE detector on \textit{Insight-HXMT}. 
 In Figure~\ref{Figurecombin}, we show results of fitting HE light curves as well as ME and LE light curves with SHO kernel.
 The three light curves are fitted well with non-white-noise structure in the ACF results and reasonable residual distributions with proper chi-square values. 
 The SHO kernel can capture two peaks associated with FRB signals in the HE light curve.
 The marginal two-peak structure also appears in fitting result ofthe  ME light curve.
In Figure~\ref{fig:2022}, we show the SHO fitting results for three bursts associated with transition state of radio bursts. The SHO parameters obtained in the fittings are given in Table ~\ref{tab:Table2}.


\subsection{PSD and Timescales}

\begin{figure}
    \includegraphics[width=0.3\textwidth]{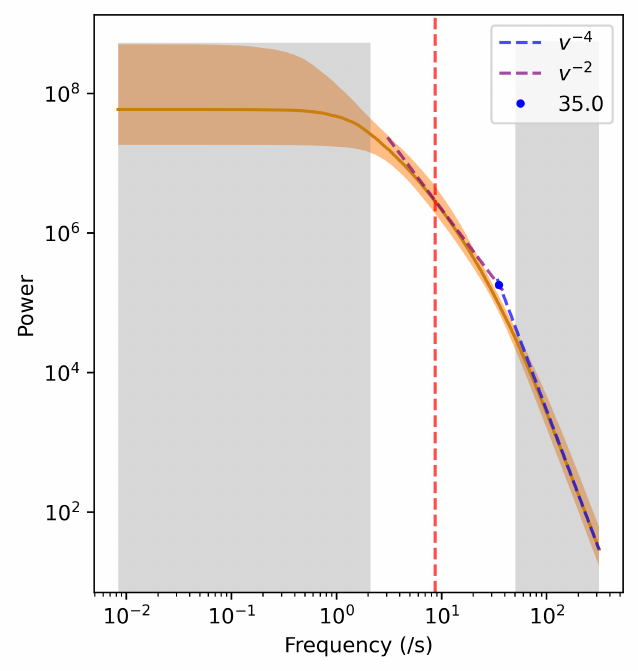}
    \includegraphics[width=0.3\textwidth]{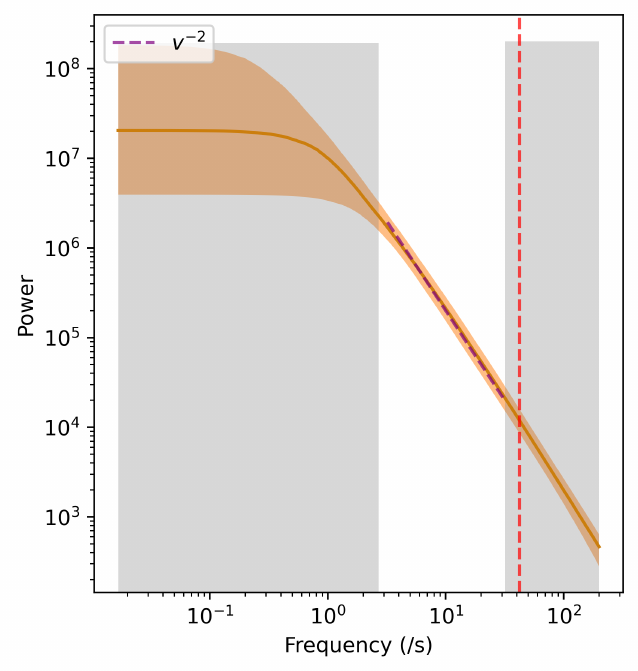}
    \includegraphics[width=0.3\textwidth]{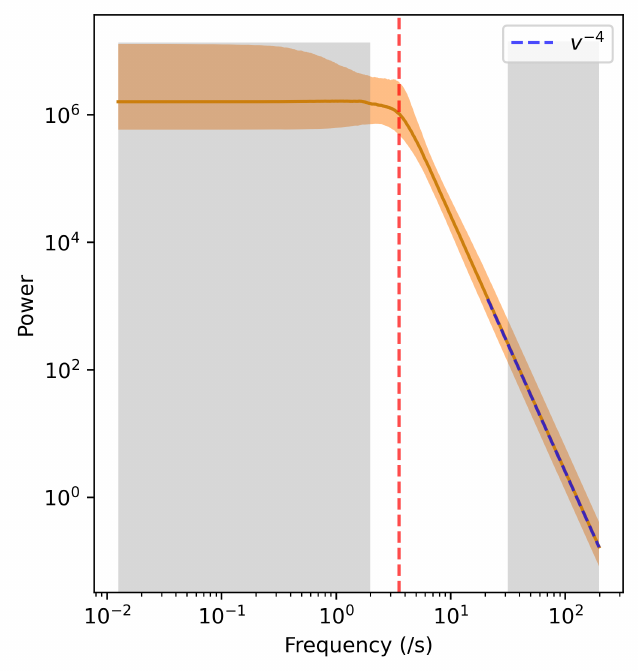}\\
    \includegraphics[width=0.3\textwidth]{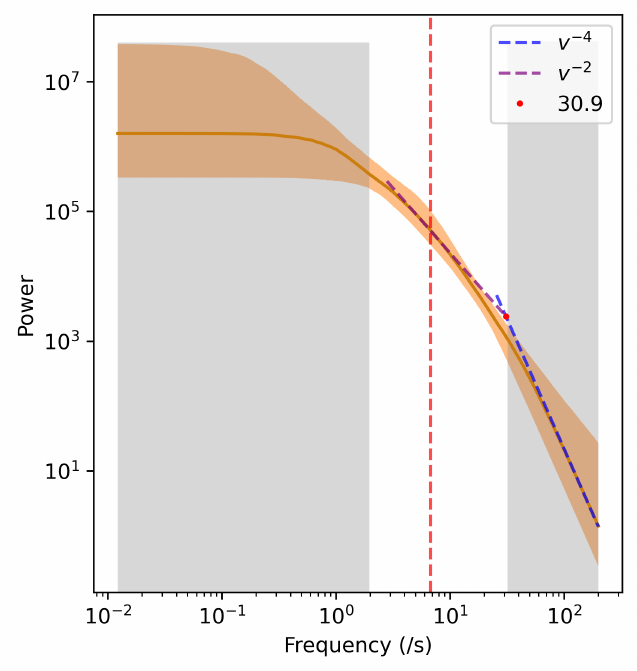}
    \includegraphics[width=0.3\textwidth]{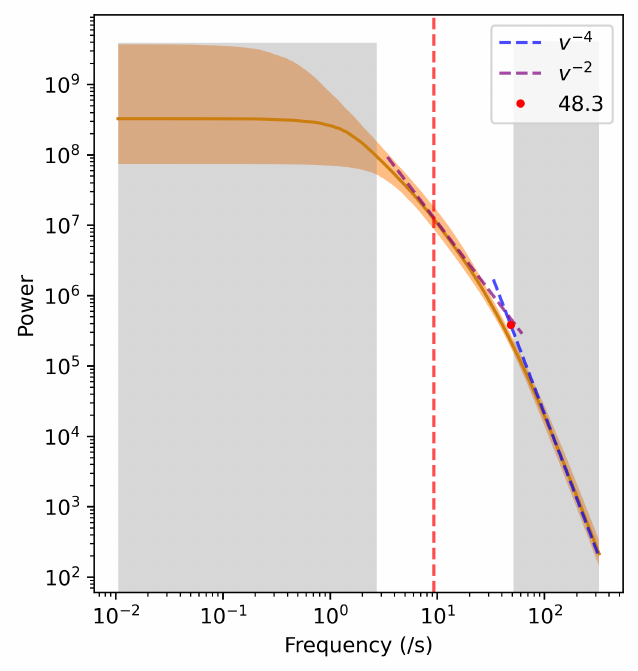}
    \includegraphics[width=0.3\textwidth]{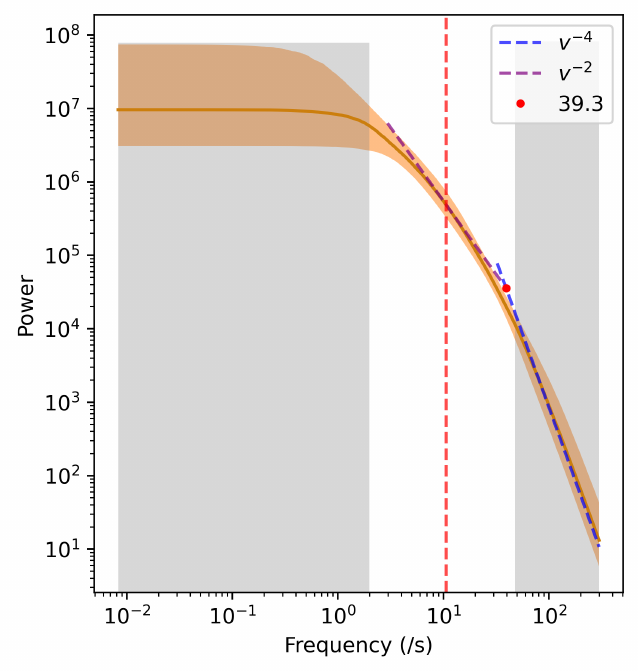}\\
    \includegraphics[width=0.3\textwidth]{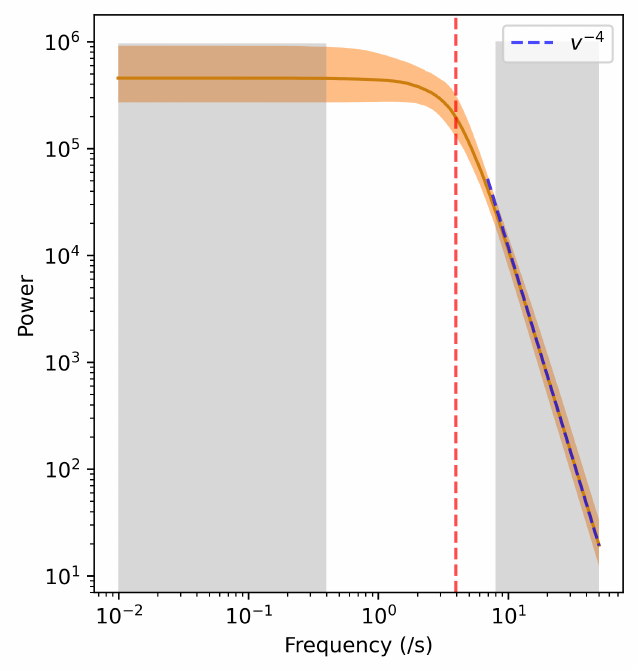}
    \includegraphics[width=0.3\textwidth]{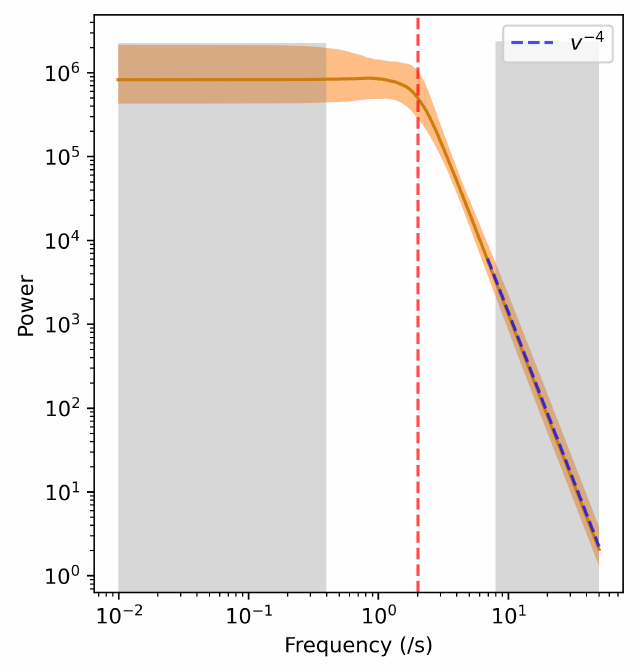}
    \includegraphics[width=0.3\textwidth]{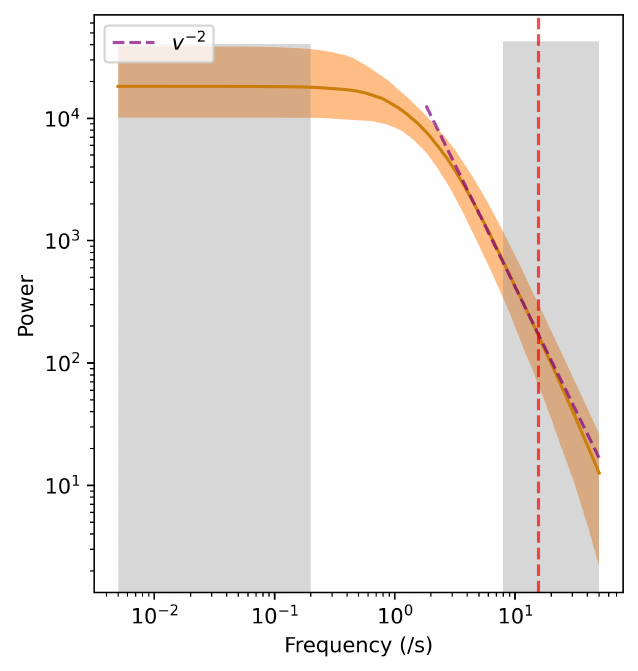}\\
    \caption{SHO PSDs derived from the fitting results.
    The PSDs in the first line are for the {\it HXMT}-HE/ME/LE light curves associated with the FRB.
    The middle three panels show the SHO PSDs extracted from the fittings of the light curves of $2020-04-29T11:13:57.650 $ (LE), NICER-2, and NICER-6.
    The bottom panels are the SHO PSDs extracted from the fittings of the GECAM light curves observed on October 14th, November 20th, 2022, and the {\it HXMT}-HE light curve on October 21st.}
    \label{psdhe}
\end{figure}  

Theoretically, there are two timescales in the SHO model \citep{1945RvMP...17..323W,2017MNRAS.470.3027K}.
In SHO PSD, the high broken frequency where spectral index shifts from $\nu^{-4}$ to $\nu^{-2}$
corresponds to a short timescale; 
and at low frequencies, the PSD changing  
from a type of red noise with the index of $\nu^{-2}$ to white noise with the index of $\nu^{0}$, 
this broken frequency corresponds to a long timescale.
However, it is difficult to obtain the two timescales simultaneously in practice,
which requires a good enough sampling rate of light curve over a long enough time.

We further build corresponding PSDs from fitting results and derive the timescales.
These timescales obtained in data are useful for investigating competition between energy release process (i.e., the disturbance) and energy dissipation (damping effect) in the XRBs.
In Figure~\ref{psdhe}, we give SHO PSDs constructed from fittings.
In each PSD figure, the two grey areas constrain valuable spectrum, 
which is defined by half of data length and one over bin size. 
The red dashed line presents $\omega_{0}$.

For the HE light curve observed by \textit{Insight}-HXMT, 
the high broken frequency is constrained at 35 Hz.
This frequency is reliable since it does not drop in the grown shadows.
For ME light curve, SHO PSD is constrained in the red noise range.
The SHO PSD for LE light curve changes from $\nu^{-4}$ directly to $\nu^{0}$.
This means that the two broken frequencies expected in SHO kernel are very close, 
approximately at 3 Hz (see the last panel in the first row of Figure~\ref{psdhe}).
We also show three SHO PSDs constructed from fitting results of one LE and two {\it NICER} light curves, 
because a possible broken frequency appears at high frequencies in three PSDs.
Unfortunately, these broken frequencies are at the edge of the shadow set by time bin of light curves.
We reanalyze the three light curves with a shorter time bin to check the reliability of the broken frequencies.
However, the broken frequencies obtained from {\it HXMT-LE} and {\it NICER-2} 
light curves are still at the edge of the shadow set by shorter time bin.
For the {\it NICER-6} light curve, the broken frequency at 39 Hz is reliable.
For the three light curves associated with transition states of radio bursts, SHO PSDs for two GECAM light curves change from $\nu^{-4}$ directly to $\nu^{0}$, 
and the broken frequency is about 2-3 Hz; 
while the PSD for HE light curve change from $\nu^{-2}$ directly to $\nu^{0}$, 
and the broken frequency is about 1 Hz.
The corresponding PSD information is given in Table~\ref{tab:Table2}.

Here, we summarize the reliable broken frequencies and corresponding PSDs for several interesting bursts.
The HE light curve is associated with FRB signals, 
and we obtain a broken frequency at 35 Hz where the index of PSD changes from -4 to -2 (Figure \ref{psdhe}).
There is no reliable characteristic frequency in SHO PSD obtained from ME light.
For the LE light, we obtain a broken frequency at 3 Hz where the index of PSD changes from -4 to 0.
On October 14, 2022, during the transition state of a radio burst, 
the SHO PSD of the GECAM light curve shows a break at 3 Hz where the index of PSD changes from -4 to 0.
Seven days later, the SHO PSD for HE light curve associated with the transient state of radio burst again shows a broken frequency at a smaller frequency of 1 Hz where the index of PSD changes from -2 to 0. 
On November 20th, GECAM detected another burst associated with a transition state of radio burst, 
and the SHO PSD of light curve also shows a break at 2 Hz where the index of PSD changes from -4 to 0.

\begin{table*}
\centering
  \begin{tabular}{l l c c c c c c c c c l}
  \toprule
  \footnotesize
  & & & & & \multicolumn{3}{c}{\small SHO Modeling Results} & \multicolumn{2}{c}{\small PSD information}\\
  \cmidrule(lr){6-8}
  \cmidrule(lr){9-10}
  & \textit{Name} & \textit{Model} & \textit{AIC} & $\chi^2$ &ln $ S_{0}$ & ln $Q$ & ln $\omega_{0}$ &\rm{Index} & \rm{broken frequency }\\
  & & & & & & & & &\rm{(Hz)}\\
  \midrule
  \footnotesize
  &NICER-1 & SHO & $-2474.083$ & 0.445 & $18.35_{-0.92}^{+1.28}$ & $-1.08_{-0.44}^{+0.39}$ & $3.91_{-0.32}^{+0.22}$ &-4&···\\
  &NICER-2 & SHO & $-1740.411$ & 0.304 & $18.58_{-1.02}^{+1.45}$& $-1.44_{-0.45}^{+0.41}$ &$4.10_{-0.37}^{+0.26}$&-2 &···\\
  & NICER-3  & SHO &$-2003.314$ &0.950& $15.62_{-0.74}^{+1.05}$ & $-1.73_{-0.98}^{+1.33}$& $3.46_{-0.25}^{+0.17}$&-4$\rightarrow$0 & 5\\
  & NICER-4  & SHO &$-1165.866$ &0.952& $16.33_{-1.18}^{+1.91}$ & $-0.24_{-0.95}^{+1.10}$& $3.44_{-0.42}^{+0.35}$&-4$\rightarrow$0& 5\\
  & NICER-5  & SHO &$-1527.372$ &0.635& $15.00_{-0.79}^{+1.32}$ & $-0.73_{-0.56}^{+0.51}$& $3.83_{-0.33}^{+0.23}$&-4&···\\
  & NICER-6  & SHO &$-2015.461$ &0.523& $16.16_{-1.19}^{+1.83}$ & $-1.65_{-0.74}^{+0.53}$& $4.02_{-0.47}^{+0.40}$&-4$\rightarrow$-2 & 39\\
  & NICER-7  & SHO &$-1505.387$ &0.521& $13.53_{-0.6}^{+0.85}$ & $0.04_{-0.51}^{+0.57}$& $4.43_{-0.19}^{+0.15}$&-4$\rightarrow$0&10\\
  & NICER-8  & SHO &$-1933.541$ &0.619& $15.08_{-0.88}^{+1.40}$ & $-0.92_{-0.53}^{+0.49}$& $3.87_{-0.36}^{+0.27}$&-4&···\\
  & NICER-11  & SHO &$-472.519$ &0.323& $13.29_{-1.15}^{+2.28}$ & $-0.84_{-1.76}^{+1.46}$& $4.81_{-0.48}^{+1.00}$&-4&···\\
  & NICER-14  & SHO &$-847.526$ &0.661& $14.06_{-1.20}^{+2.19}$ & $-0.67_{-1.50}^{+1.10}$& $3.95_{-0.44}^{+0.53}$&-4&···\\
  &20200429T11\_13\_57.650(LE)& SHO &$-1332.795$ &0.916& $13.48_{-1.37}^{+2.18}$ & $-1.76_{-1.40}^{+1.01}$& $3.86_{-0.55}^{+1.04}$&-2&···\\
  &20200514T14\_49\_22.000(LE)& SHO &$-1080.988$ &0.985& $12.85_{-0.70}^{+0.97}$ & $-0.37_{-0.53}^{+0.58}$& $4.25_{-0.23}^{+0.23}$&-4$\rightarrow$0&10\\
  &20200508T06\_17\_16.589(LE)& SHO &$-866.955$ &0.824& $12.53_{-1.01}^{+1.88}$ & $-0.91_{-1.18}^{+0.68}$& $4.37_{-0.40}^{+0.49}$&-4$\rightarrow$0&9\\
  &20200520T14\_10\_49.780(LE)& SHO &$-1068.439$ &0.878& $14.32_{-1.35}^{+2.47}$ & $-2.02_{-1.54}^{+0.84}$& $4.33_{-0.67}^{+0.98}$&-2&···\\
  &X-gamma co-emitting ray(LE)& SHO &$-3324.523$ &0.889& $14.18_{-0.80}^{+1.33}$ & $-0.65_{-0.56}^{+0.62}$& $2.84_{-0.30}^{+0.20}$&-4&···\\
  &20200428T09\_51\_04.634(ME)& SHO &$-2498.810$ &1.084& $12.47_{-0.62}^{+0.90}$ & $-0.76_{-0.66}^{+1.01}$& $2.55_{-0.19}^{+0.13}$&-4$\rightarrow$0& 2\\
  &20200503T23\_25\_13.250(ME)& SHO &$-1076.587$ &0.786& $14.91_{-1.13}^{+1.92}$ & $-0.64_{-0.73}^{+0.79}$& $3.53_{-0.44}^{+0.30}$&-4&···\\
  &20200510T06\_12\_01.622(ME)& SHO &$-1801.796$ &0.536& $16.50_{-1.35}^{+1.83}$ & $-1.09_{-0.59}^{+0.79}$& $3.53_{-0.44}^{+0.30}$&-4&···\\
  &20200510T21\_51\_16.221(ME)& SHO &$-1813.762$ &0.850& $14.81_{-0.97}^{+1.71}$ & $-0.53_{-0.67}^{+0.57}$& $3.58_{-0.36}^{+0.24}$&-4$\rightarrow$0&6\\
  &20200520T14\_10\_49.780(ME)& SHO &$-614.023$ &1.119& $9.76_{-0.91}^{+1.62}$ & $-0.54_{-0.85}^{+1.09}$& $3.56_{-0.33}^{+0.22}$&-4$\rightarrow$0&5\\
  &20200520T21\_47\_07.480(ME)& SHO &$-1188.322$ &0.731& $13.47_{-0.98}^{+1.78}$ & $-0.59_{-0.62}^{+0.53}$& $3.77_{-0.45}^{+0.24}$&-4&···\\
  &X-gamma co-emitting ray(ME)& SHO &$-3170.246$ &1.631& $15.62_{-1.13}^{+2.76}$ & $-0.91_{-1.41}^{+0.79}$& $2.84_{-0.45}^{+0.37}$&-2&···\\
  &20200428T08\_05\_50.080(HE)& SHO &$-1245.256$ &0.642& $11.44_{-1.07}^{+2.02}$ & $-0.85_{-0.91}^{+0.73}$& $3.84_{-0.44}^{+0.30}$&-4&···\\
  &20200510T21\_51\_16.221-1(HE)& SHO &$-1455.960$ &0.710& $12.85_{-0.57}^{+0.68}$ & $-0.08_{-0.42}^{+0.53}$& $4.42_{-0.16}^{+0.14}$&-4$\rightarrow$0&10\\
  &20200510T21\_51\_16.221-2(HE)& SHO &$-1819.018$ &0.829& $11.22_{-0.60}^{+0.92}$ & $-0.74_{-0.49}^{+0.40}$& $4.35_{-0.24}^{+0.20}$&-4&···\\
  &20200510T21\_51\_16.221(HE)& SHO &$-1455.960$ &0.710& $12.85_{-0.57}^{+0.68}$ & $-0.08_{-0.42}^{+0.53}$& $4.42_{-0.16}^{+0.14}$&-4&···\\
  &P051435700104-20221013-01-01(HE)& SHO &$-1129.101$ &0.712& $8.91_{-0.93}^{+1.74}$ & $-0.35_{-1.31}^{+0.97}$& $4.05_{-0.30}^{+0.48}$&-4&···\\
  &P051435700106-20221013-01-01(HE)& SHO &$-1016.236$ &1.096& $11.18_{-0.97}^{+1.56}$ & $-0.37_{-1.04}^{+0.95}$& $3.79_{-0.37}^{+0.40}$&-4&···\\
  &P051435700109-20221014-02-01(HE)-1& SHO &$-3442.133$ &1.016& $14.51_{-0.52}^{+0.65}$ & $-0.46_{-0.30}^{+0.28}$& $3.78_{-0.17}^{+0.14}$&-4$\rightarrow$0&7\\
  &P051435700109-20221014-02-01(HE)-2& SHO &$-3020.944$ &0.638& $14.61_{-0.51}^{+0.66}$ & $-0.64_{-0.28}^{+0.25}$& $4.20_{-0.16}^{+0.13}$&-4$\rightarrow$0&10\\
  &P051435700110-20221014-02-01(HE)-1& SHO &$-1619.972$ &0.611& $14.29_{-0.73}^{+1.22}$ & $-0.70_{-0.44}^{+0.42}$& $4.00_{-0.27}^{+0.19}$&-4&···\\
  &P051435700110-20221014-02-01(HE)-2& SHO &$-1043.588$ &0.578& $12.72_{-0.64}^{+0.92}$ & $-0.39_{-0.46}^{+0.47}$& $4.26_{-0.23}^{+0.20}$&-4$\rightarrow$0&9\\
  &P051435700608-20221024-02-01(HE)& SHO &$-1794.783$ &0.628& $9.78_{-0.50}^{+0.68}$ & $-0.70_{-0.54}^{+0.39}$& $4.95_{-0.19}^{+0.26}$&-4$\rightarrow$0&11\\
  &P051435700609-20221024-02-01(HE)& SHO &$-801.477$ &0.504& $10.68_{-1.00}^{+1.71}$ & $-0.72_{-0.84}^{+0.67}$& $4.46_{-0.40}^{+0.37}$&-4&··· \\
  &P051435700610-20221024-02-01(HE)& SHO &$-1199.093$ &0.748& $8.50_{-0.60}^{+0.95}$ & $-0.02_{-0.99}^{+0.64}$& $4.26_{-0.18}^{+0.31}$&-4$\rightarrow$0& 10\\
  &P051435700617-20221025-03-01(HE)& SHO &$-1182.540$ &0.549& $10.26_{-0.83}^{+1.32}$ & $-0.98_{-1.47}^{+0.66}$& $4.39_{-0.36}^{+0.90}$&-4 &··· \\
  &X-gamma co-emitting ray(HE)& SHO &$-2346.845$ &0.755& $17.37_{-1.14}^{+2.30}$ & $-1.49_{-0.69}^{+0.49}$& $3.95_{-0.55}^{+0.33}$&-4$\rightarrow$-2& 35\\
  &TransitionFRB20221021(HE)& SHO &$-3057.224$ &0.842& $9.30_{-0.58}^{+0.70}$ & $-2.27_{-0.89}^{+1.01}$& $4.60_{-1.09}^{+0.81}$&-2$\rightarrow$0&2\\
  &TransitionFRB20221014(GECAM-B)& SHO &$-1713.131$ &0.493& $12.56_{-0.47}^{+0.58}$ & $-0.43_{-0.38}^{+0.34}$& $3.22_{-0.16}^{+0.16}$&-4$\rightarrow$0&4\\
  &TransitionFRB20221120(GECAM-B)& SHO &$-1640.944$ &0.637& $13.11_{-0.64}^{+0.97}$ & $-0.22_{-0.50}^{+0.50}$& $2.53_{-0.21}^{+0.17}$&-4$\rightarrow$0&2\\
  \bottomrule
\end{tabular}
\caption{SHO model fitting outcomes and the derived PSD information. For objects with broken frequency value in PSD information, the index refers to the  index value  above the broken frequency to that  below the  broken frequency, and for objects without broken frequency, the PSD is a single power-law form in the credible interval
}
\label{tab:Table2}
\end{table*}

\section{Discussion}\label{sec:discussion}
We can build a physics scenario to understand the fitting results, especially the timescales derived in SHO PSDs.
When a system is disturbed by energy release, it will run out of a steady state.
There must be a responding or 
retarding time that is determined by the characteristics of system and disturbance.
In this stage, the system responds weakly to the disturbance (corresponding to the regime of $\nu^{-4}$ in the SHO PSD).
After this timescale, the state of the system will vary significantly due to the disturbance of injection of energy.
Meanwhile, there are some kinds of energy loss processes existing in system.
After a period of time, system will achieve equilibrium when the energy net-exchange is zero.
Note that there must still exist fluctuations around the equilibrium.

Looking back at the SHO PSD, 
the short timescale corresponding to the broken frequency where the PSD index changes from -4 to -2 can be considered as the responding/retarding timescale; the long timescale corresponding to the broken frequency where the PSD index changes from -2 to 0 can be considered as the equilibrium/relax timescale.

The {\it HMXT}-HE burst associated with the FRB is special in our analysis.
A clear retarding timescale is constrained from this burst.
This timescale 
is related to the interplay between inertia of system and disturbance acting on the system.
This indicates that HE photons are radiated nearly simultaneously with the disturbance process.
The SHO PSD of ME light curve is in the region of red noise. This implies that ME photons are produced after responding to the system and before equilibrium.
The HE and ME photons may be produced by the same emission mechanism, 
but ME photons are produced in a larger region.

For LE burst, SHO PSD changes from $\nu^{-4}$ to $\nu^{0}$ directly.
This can be explained as the retarding timescale is very close to the equilibrium timescale.
This situation is different from that of HE and ME bursts. 
The LE burst could be a normal X-ray burst, and is irrelevant to the production of HE-ME and FRB photons.

The retarding timescale obtained from the HE burst is about 0.03 s, corresponding to the broken frequency of 35 Hz.
This implies that the timescale of energy disturbance $t_{\rm inj}$ should not be longer than 0.03 s, i.e., $t_{\rm inj}\leq0.03$ s.

The other special burst is the {\it NICER}-6.
The retarding timescale is also constrained by light curve as $\approx1/(39 {\rm\ Hz})\approx0.02$ s.
This is similar to {\it HXMT}-HE burst associated with FRB, but in the lower X-ray energies.
If our discussion on HE burst associated with FRB 200428 is general, 
the X-ray burst of {\it NICER}-6 could have associated radio pulses.

The X-ray bursts observed by GECAM and {\it HXMT}-HE between October 2022 and November 2022 
were considered to be associated with the transition states to FRB events.
In our analysis, the bursts do not show evidence of the retarding timescale.
A long equilibrium timescale of $\approx0.3-1$ s is obtained.
A similar result is also found in the NICER light curve.

The proposed scenario can be used to understand the evolution of the XRBs of magnetars.
These timescales we obtained can put constraints on the production mechanisms of XRBs.
In particular, we would like to stress that our analysis of the {\it HXMT}-HE burst associated with FRB 200428 
suggests a relevant energy disturbance timescale of $t_{\rm inj}\leq0.03$ s, and the HE photons are produced quasi-simultaneously with the response to the energy disturbance.

The timescale smaller than 0.03 s we got is similar to the magnetic reconnection time in \citet{2021ApJ...921...92B}. 
As shown in the light curve profiles, both the LE and ME light curves have extra hills before the main peaks, 
which might indicate lower energy photons are emitted in the magnetic line approaching the reconnecting threshold. 
While in this process, higher energy photons uneasily escape with strong interaction
and later are released nearly at the same time as the FRB emitting. 

\section{Acknowledgement}
We thank the anonymous referee for the valuable comments that have helped improve the paper. This work made use of data from the Insight-HXMT mission, a project funded by China National Space Administration and the Chinese Academy of Sciences. We acknowledge funding support from the National Natural Science Foundation of China (NSFC) under grant Nos. 12393852, 12122306 and 12333007, as well as support from the Strategic Priority Research Program of the Chinese Academy of Sciences (No. XDB0550300).

\newpage
\appendix
\begin{figure}[H]
    \includegraphics[width=0.3\textwidth]{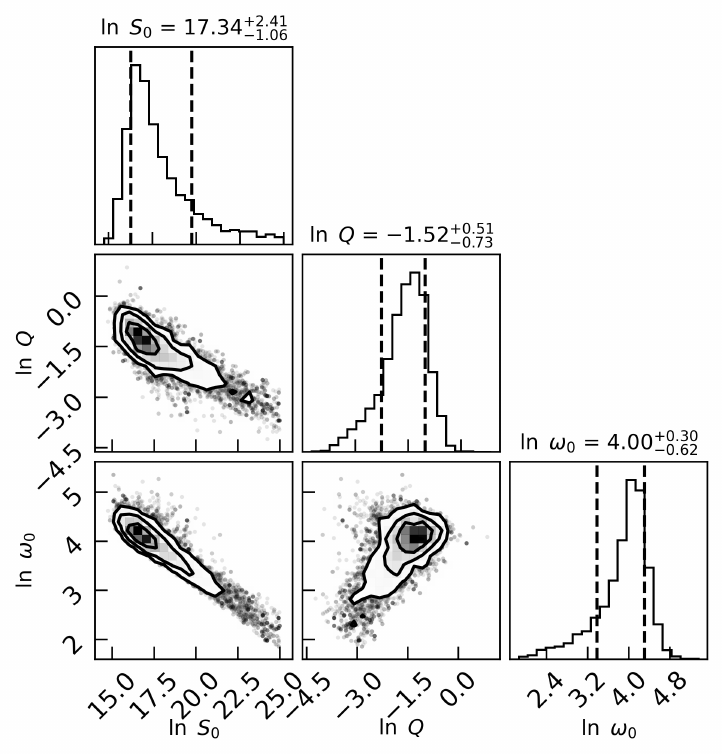}
    \includegraphics[width=0.3\textwidth]{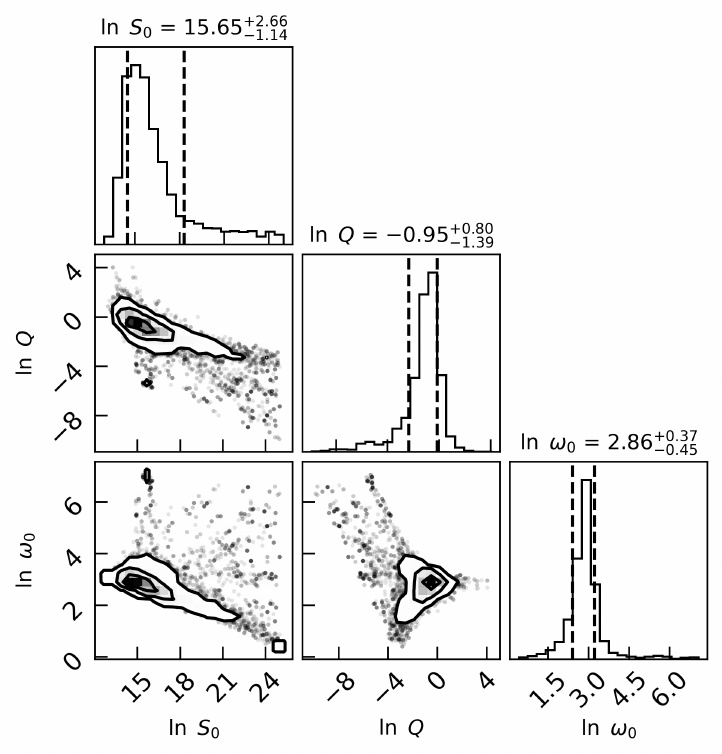}
    \includegraphics[width=0.3\textwidth]{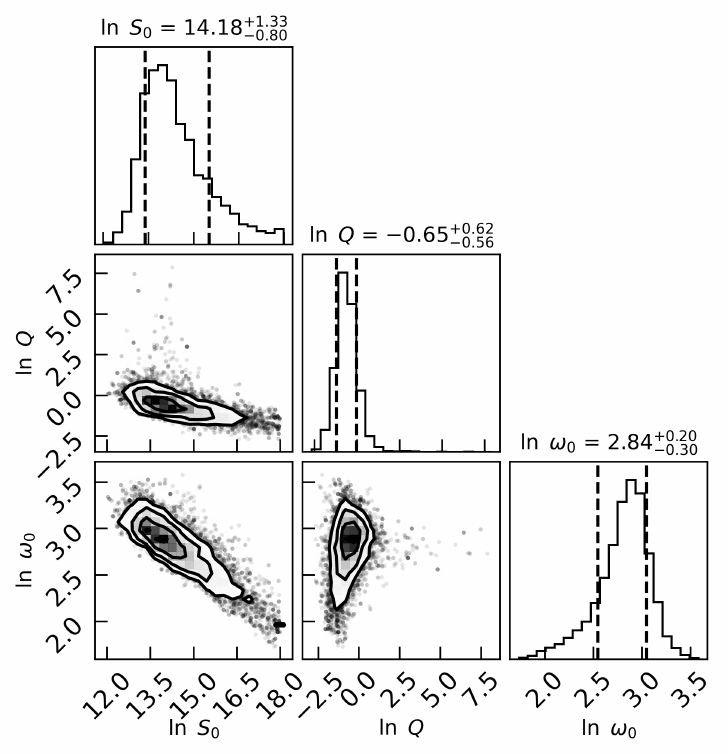}\\
    \includegraphics[width=0.3\textwidth]{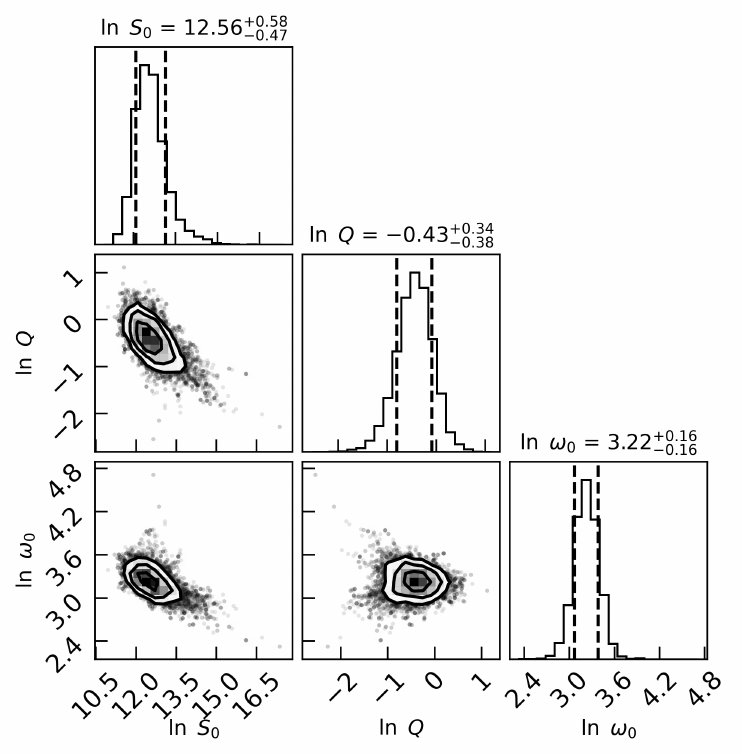}
    \includegraphics[width=0.3\textwidth]{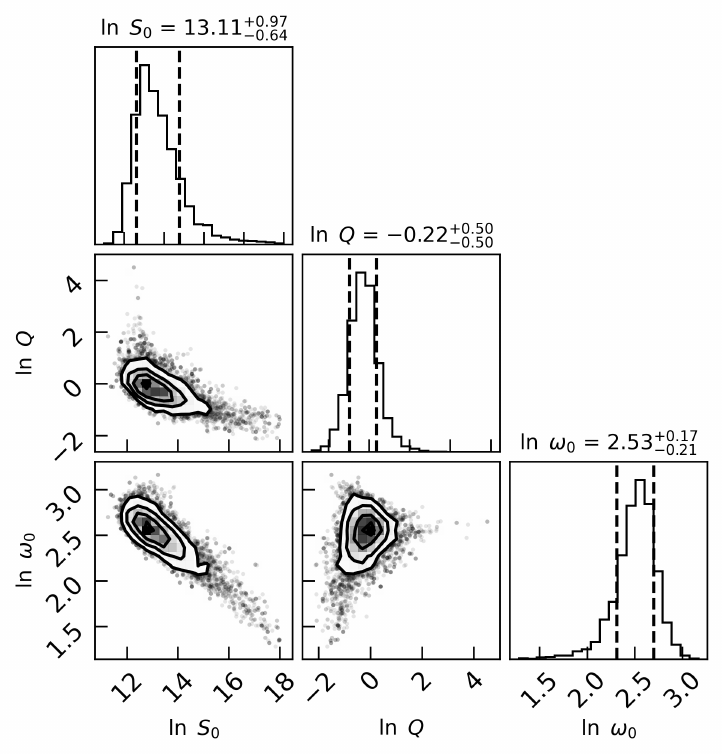}
    \includegraphics[width=0.3\textwidth]{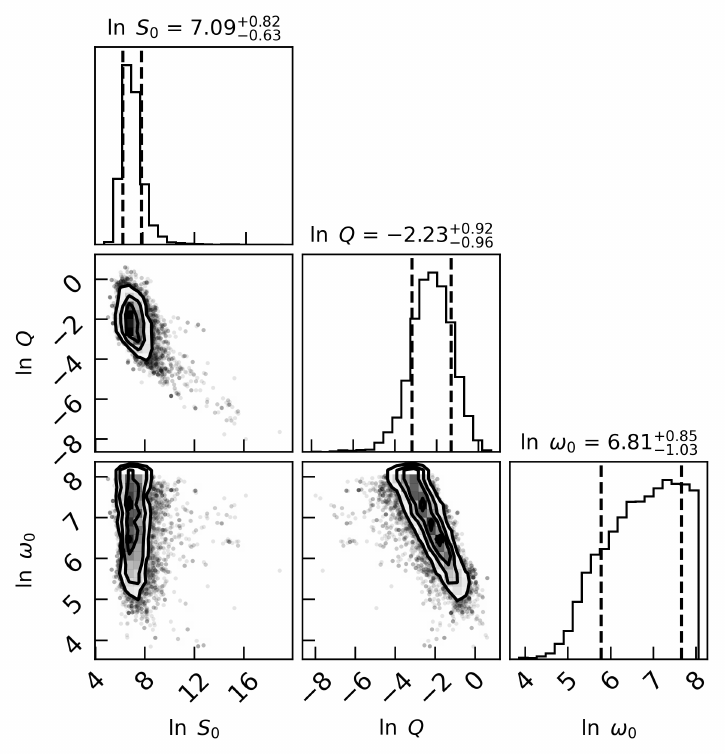}\\
    \includegraphics[width=0.3\textwidth]{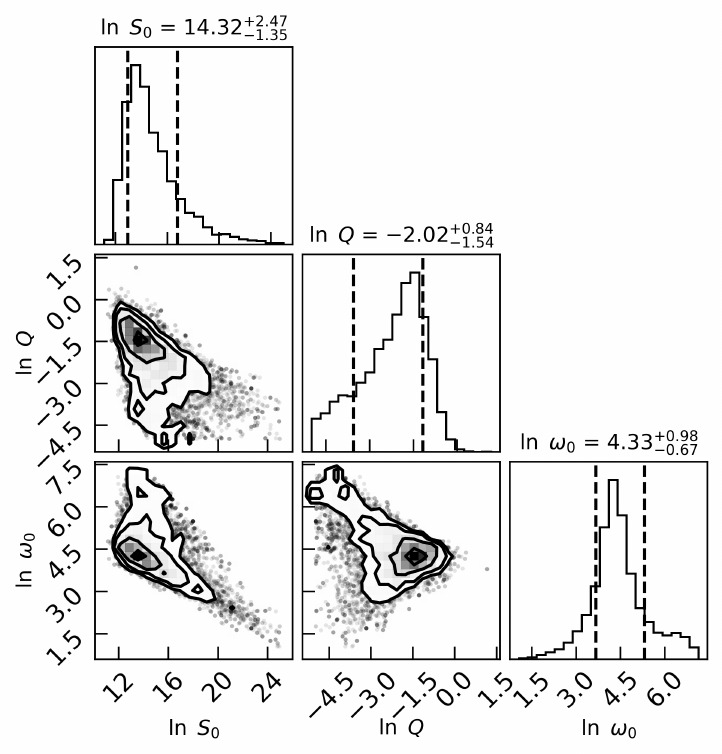}
    \includegraphics[width=0.3\textwidth]{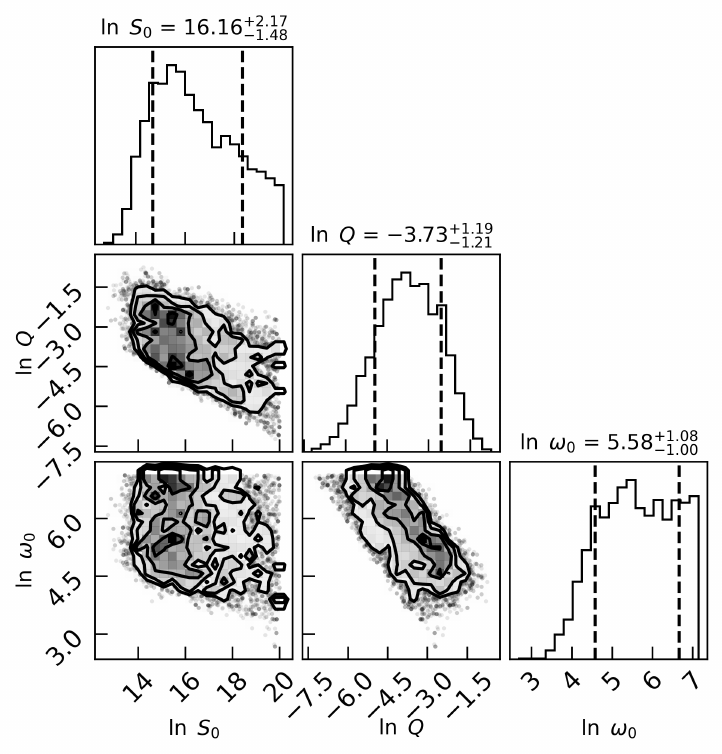}
    \includegraphics[width=0.3\textwidth]{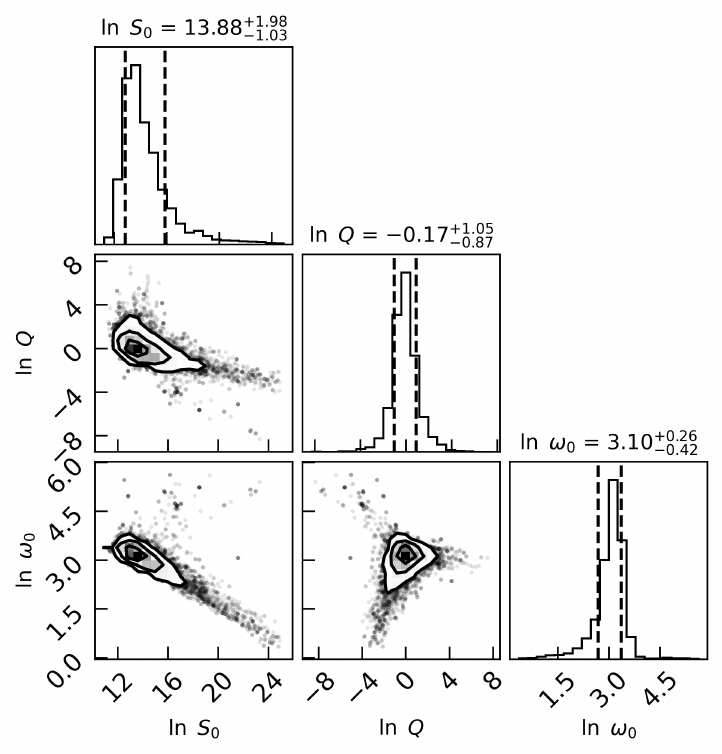}
    \caption{Posterior probability densities of model parameters for SHO. For the top three panels, simulating results from the X-ray co-emitting data observed by the HE, ME, and LE detectors respectively. The middle three panels show the other three corner plots extracted from bursts on Oct. 14th and Nov. 20th detected by GECAM-B and Oct. 21st detected by HXMT. The bottom three panels are corner plots from 2020-04-29 11:13:57.650 (LE), and cut ME and LE for co-emitting events respectively. The median value and 68\% confidence intervals of the parameter distribution are marked by the vertical dotted lines.}
    \label{parameter-pdf}
\end{figure}

\bibliography{sample631}
\bibliographystyle{aasjournal}



\end{document}